\documentclass{ws-p8-50x6-00}

\newcommand{\ba}{\begin{eqnarray}}
\newcommand{\ea}{\end{eqnarray}}

\begin{document}

\title{Report of the Working Group on Goldstone Bosons\lowercase{$^x$}}

\author{Johan Bijnens\footnote{C\lowercase{onveners}.~~~~ $^\dagger$
P\lowercase{articipants, the institutes are given in the individual
contributions}.}}

\address{Dept. of Theoretical Physics 2, Lund University,\\
S\"olvegatan 14A, S22362 Lund, Sweden}

\author{Ada Farilla$^*$}

\address{INFN - Sezione Roma3, Rome, Italy}

\author{Rory Miskimen$^*$}

\address{Dept. of Physics, University of Massachusetts,\\
 Amherst, MA 01003, USA }

\author{F.~Ambrosino$^\dagger$, M.~Arenton$^\dagger$, P.~Cenci$^\dagger$,
V.~Cirigliano$^\dagger$, A.~Fariborz$^\dagger$,
A.~Gasparian$^\dagger$, M.~Golterman$^\dagger$, R.~Kaiser$^\dagger$,
D.~Mack$^\dagger$, B.~Moussallam$^\dagger$,
T.~Nakano$^\dagger$, B.~Nefkens$^\dagger$, A.~Nyffeler$^\dagger$,
J.~Oller$^\dagger$,
E.~Oset$^\dagger$, J.R.~Pelaez$^\dagger$, J.~Palomar$^\dagger$,
A.~Radyushkin$^\dagger$, P.~Rubin$^\dagger$, J.~S\'a~Borges$^\dagger$,
J.~Schacher$^\dagger$, S.~Schmidt$^\dagger$, J.~Stern$^\dagger$,
T.~Walcher$^\dagger$}


\maketitle

\abstracts{An overview is presented of the talks in the working
group on Goldstone Bosons. Topics touched on are CP-violation
in the Kaon system, rare Kaon decays, $\pi\pi$-scattering, $\phi$-meson
decays, scalar mesons, form-factors and polarizabilities, $\eta$-decays,
chiral symmetry breaking, connections with QCD at short-distances
and effective theories for electroweak physics.
}

\footnotetext{
$^x$Talk presented at Chiral Dynamics 2000, Jefferson Lab, Newport News,
Virginia, USA, July 17-22, 2000.}
\vskip-13cm\phantom{p}\hfill LU TP 00-50\\\phantom{p}\hfill hep-ph/0012024\\
\phantom{p}\hfill November 2000\\[11cm]

\enlargethispage{1cm}

\section{Introduction}

In this talk we report on the working group on Goldstone Bosons
of the Chiral Dynamics 2000, theory and Experiment,
meeting at Jefferson Laboratory, Newport News,
Virginia, USA, July 17-22, 2000.
We had 28 presentations during four afternoon sessions  divided
into 7 main areas. There was of course the unavoidable overlap
between the different main areas and with the  plenary talks. 
In this report we summarize the main
issues discussed in our working group. More details for each of the
presentations and further references
can be found in the individual contributions which follow
this overview. The discussion below and the order of the contributions 
is organized by the chronological order of the presentations
at the meeting.

\section{Kaons}

Kaon-physics covered a little more than one of the afternoons
in our working group. In addition there were four plenary talks
that can be classified as belonging to this part.

The discussions were divided in three areas: $K\to\pi\pi$-decays
and $CP$-violation, rare Kaon decays and the third topic
was specifically related more to $\pi\pi$-scattering, i.e. the
calculations and measurements of $K_{\ell4}$-decays.

\subsection{Rare Kaon Decays}

The subject of rare Kaon decays was covered in talks by
Nakano\cite{Nakano}, Schmidt\cite{Schmidt} and Arenton.
In addition Isidori gave an overview in his plenary talk\cite{Isidori}.

The first set of rare decays discussed fall under the category tests
of chiral dynamics. Here we have the new BNL result on
$K^+\to\pi^+\pi^0\gamma$ on the direct photon emission
contribution\cite{Nakano}. This decay is interesting since its
magnetic amplitude tests the interplay of the anomaly and the weak interaction.
The result is somewhat surprising in that it indicates no
contribution from the direct weak-anomaly term\cite{BEP}.
The second set of weak CHPT-testing decays on which new data were
presented was $K_L\to\pi^0\gamma\gamma$ and $K_S\to\gamma\gamma$.
The new NA48 data on the latter decay\cite{Schmidt,NA48KS} are in
perfect agreement with the older data and with the parameter-free
CHPT prediction\cite{KSgg}. In $K_L\to\pi^0\gamma\gamma$
a similar parameter-free predictions at order $p^4$ exists\cite{KLpgg}
but while it predicts the shape of the di-photon mass spectrum well,
it does underestimate the rate. The new data presented here show now
clear evidence for the vector-meson-exchange contribution at low values
of $m_{\gamma\gamma}$. This will hopefully allow for a clean determination
of both the $p^4$ CHPT effects and the additional $p^6$ contributions.

The question of Vector-Meson-Exchange contributions is also important in
the rare decays presented by Arenton\cite{Arenton} as discussed
in e.g. Ref.\cite{eeee}. The knowledge of all the decay modes of this type
allows to disentangle the various models. The decay $K_L\to\pi^+\pi^-e^+e^-$
has a $T$-violating asymmetry in the final state that is observed
in agreement with predictions\cite{kppee}. Further refinement of the
latter mode might allow a test of the chiral prediction for the neutral
Kaon charge radius.

\subsection{$CP$-violation and $K\to\pi\pi$-decays}

This area was covered experimentally and theoretically. The experimental
results of KTeV and NA48 were treated by Arenton\cite{Arenton,KTeV}
and Cenci\cite{Cenci,NA48} respectively. The NA48 results
are combined from the 97 and 98 data. The KTeV results are from 20\% of the
96/97 data. Both experiments are in the process of analyzing substantially
larger data sets. So we hope that the difference
\ba
{\mbox{Re}}({\varepsilon^\prime}/{\varepsilon})(\mbox{NA48}) &=&
(14.0\pm4.3)\times 10^{-4}\nonumber\\
\mbox{Re}({\varepsilon^\prime}/{\varepsilon})(\mbox{KTeV}) &=&
(28.0\pm4.1)\times 10^{-4}
\ea
will soon be resolved\footnote{Systematic and statistical errors were
combined quadratically.}. 
These results still present a clear indication that the source of
$CP$-violation is at the weak scale and not at some very much higher
scale. In that sense the qualitative prediction of Gilman
and Wise\cite{GilmanWise} has been confirmed.

The measurement at DA$\Phi$NE should provide an independent confirmation
of this result using an entirely different approach. This was discussed
in the plenary talk by P.~Franzini\cite{Franzini}.

On the theory side, things have progressed but are still rather uncertain.
The problems facing lattice gauge theory calculations but also the way
to solutions, which involves a lot of chiral dynamics,
were discussed in some detail by M.~Golterman\cite{GoltermanPlenary}.
A short overview of the theoretical problems facing analytical
approaches was presented by Isidori in his plenary talk.
An optimistic view of the results of analytical calculations was
presented by Bijnens\cite{Bijnenseps}, where it was concluded
that both the $\Delta I=1/2$ rule and the above given values
of $\varepsilon^\prime/\varepsilon$ can be understood in the
framework of the $1/N_c$-expansion enhanced with modeling the intermediate
energy regime.
References to other approaches can be found in Ref.\cite{Isidori}
or in the lectures Ref.\cite{Benasque}.

That  $\varepsilon^\prime/\varepsilon$ is a quantity beset by small
effects was exemplified by Cirigliano\cite{Cirigliano} who showed
that electromagnetic effects beyond the electromagnetic
Penguin\cite{EMpenguin} can play a rather important role. In particular
the discrepancy between the phase deduced from $\pi\pi$-scattering
and the phase in $K\to\pi\pi$ could come from this source.
A similar analysis can be found in Ref.\cite{EIMNP}.

\subsection{$K_{\ell4}$}

This decay was discussed in two contributions at this meeting.
There are the new data from BNL\cite{Zeller} on $K^+\to\pi^+\pi^-e^+\nu$.
These show the future precision to be expected in fixed target
measurements of $\pi\pi$-scattering and on the three-form-factors
and their dependence on the kinematical variables.
In anticipation of these new measurements the old Pais-Treiman method has
been reanalyzed\cite{PT} and the alternative method of using form-factor
parametrizations was updated to reflect the expected precision in
Ref.\cite{AB}.

The other talk concerned the theoretical progress on this
decay\cite{Bijnenskl4}.
Since the absolute values of the form-factors
in $K_{\ell4}$ are the major input for $L_1$, $L_2$ and $L_3$
an analysis beyond the existing one-loop and dispersive ones in CHPT\cite{KL4}
became necessary. The calculation does now exist to two-loops\cite{ABT} and
was fitted to the existing data\cite{Rosselet}. The new data from
BNL and Frascati will improve the precision rather
much. The predictions for $\pi\pi$-scattering using these parameters
also agree well with the data and the Roy analysis\cite{Bijnenskl4,Colangelo}

\section{$\mathbf \pi\pi$-Scattering}

The subject of $\pi\pi$-scattering was discussed rather extensively
at this meeting. A first introduction to the CHPT calculation was given
by Leutwyler\cite{Leutwyler} and more details as well as
the results of the new Roy equation analysis\cite{Roynew}
by Colangelo\cite{Colangelo}, see also
Ref. \cite{Royold}. The theoretical issues involved were 
also discussed by Gasser in his talk about sigma terms\cite{Gasser}.

As discussed above the calculation of $K_{\ell4}$ was used to predict the
$\pi\pi$-scattering lengths and it fitted well. The other to be
expected precise measurement comes from pionium decays.
The theory behind these decays including isospin breaking and
especially photonic corrections was discussed by A.~Rusetsky\cite{Rusetsky}.
The experimental side, with the DIRAC proposal, was presented in our
working group by Schacher\cite{Schacher}. The highlight here was the
preliminary evidence for atomic pairs showing that the prospects for
an accurate measurement of $a_0^0-a_0^2$ via the pionium lifetime
are very good.

That measurements very near threshold are important to determine the
threshold parameters was also stressed by J. S\'a Borges in
his talk\cite{Saborges}. Using a simple unitarization procedure
they showed that the present $\pi\pi$-data set is not sufficient to
discriminate between the small and large condensate scenario. The latter
 is one of the main reasons for the 
theoretical and experimental effort
regarding $\pi\pi$-scattering. We will discuss this aspect below
in Sect. \ref{Other}.

\section{$\mathbf \phi$}
$\phi$ decays are a very useful ground for different aspects of
chiral dynamics and their importance has been underlined in many talks of
this working group.
\newline
J.R. Pelaez \cite{pelaez1}, in his talk on  $\phi\to\pi\pi$, has drawn our 
attention on this decay which is interesting since it violates isospin, it
is OZI suppressed and has an interplay with the $\phi-\omega$ mixing.
Moreover, the most recent measurements at the VEPP-2M collider
are in conflict: CMD-2 measures $BR=(2.20\pm0.25\pm0.20)\times 10^{-4}$
while the SND result is $BR=(0.71\pm0.11\pm0.09)\times 10^{-4}$. The main 
contributions to this process come from $\phi-\rho$ mixing and from 
$\phi-\omega-\rho$ mixing. In the first case we can distinguish the
electromagnetic contribution $\phi-\gamma-\rho-\pi\pi$ and the strong
contribution due to Kaon loops. In this latter case Pelaez and
collaborators\cite{pelaez2} have used the unitarized chiral amplitude
with strong
isospin breaking to calculate $g^{K\bar{K}}_{\phi\pi\pi}$.
For the $\phi-\omega-\rho$ mixing the three different scenarios present in
the literature have been taken into account: the ``Strong $\phi-\omega$
mixing'', the ``Weak $\phi-\omega$ mixing'' and the ``Hidden Local Symmetry
(HLS)'' scenario. The final results depend on the $\phi-\omega-\rho$
mixing: the Strong scenario gives a result compatible with the CMD-2
measurement while the Weak and HLS give results comparable with SND. New
and more precise experimental results will play a crucial role in solving
the  question. 
\vskip 0.2cm
In the work presented by E. Oset \cite{oset1} on $\phi$ radiative decays again a chiral
unitary approach is used to describe the decays $\phi\to\pi^0\pi^0\gamma$
and  $\phi\to\eta\pi^0\gamma$. These decays are forbidden at tree level but
can proceed via Kaonic loops, which involve $K^+K^-$ transition to 
$\pi^0 \pi^0$ and  $\pi^0 \eta$ where the $f_0(980)$ and $a_0(980)$ 
resonances appear. The definition of the nature of these two scalar mesons
is a longstanding problem in light meson spectroscopy and many hypothesis,
like $qq \bar{q}\bar{q}$ or $K \bar{K}$ molecules, apart from the standard
$q \bar{q}$, have been made in the last decades. In the framework
developed by Oset and collaborators\cite{oset2,oset3,oset4}
these resonances would be ordinary
meson-meson scattering resonances coming from multiple scattering of the
mesons. The diagrams of these decays involve charged Kaon loops with the
photon coupled in all possible ways to the loops and to the final state 
particles. The  $\pi^0 \pi^0$ and  $\pi^0 \eta$ mass spectrum have been
evaluated and both the $f_0$ and the $a_0$ are well visible in such
distributions and in good agreement with the experimental spectra from
VEPP-2M experiments, CMD-2 and SND. In the case   $\phi\to\eta\pi^0\gamma$
there is also a good agreement with the branching ratio calculated in the
framework of chiral unitary approach ($0.87 \times 10^{-4}$) and the
results of SND $(0.83 \pm 0.23)\times 10^{-4}$ and CMD-2   
$(0.90\pm 0.24\pm 0.10)\times 10^{-4}$, thus representing a successful
application of such approach.
\vskip 0.2cm
$\phi$ radiative decays are being studied experimentally at CEBAF as
reported by P. Rubin \cite{rubin1} in his talk on the RadPhi 
experiment (TJNAF E94-016)
approved in 1995 and which has started taking data in summer 2000. It is a
neutral apparatus suited for photo-production of the $\phi$ meson and   for
the detection of $\phi\to$all-photon final state decay modes such as  
 $\phi\to f_0(980)\gamma$,  $\phi\to a_0(980)\gamma$,  $\phi\to\rho\gamma$,
 $\phi\to\eta'\gamma$,  $\phi\to\omega\pi^0$.
The physics motivations of RadPhi are: the nature of $f_0(980)$ and
$a_0(980)$, the radiative width of the $\phi$, the gluonic and strangeness
content of $\eta'$, searches for C- and I-violating decays, the radiative
decays of the $\rho$ and $\omega$.
\newline
Similar studies have been done in the last years at the VEPP-2M collider,
by the SND \cite{rubin2} and CMD-2  \cite{rubin3} experiment, and are being 
done presently at the
DA$\Phi$NE collider by the KLOE experiment \cite{Franzini,rubin5,rubin6}.
The RadPhi apparatus is capable
of photo-producing $\sim$2.5 million $\phi$ mesons per day with a photon
beam of 5.65 GeV and all-photon final states are detected with efficiency
ranging from 3$\%$ to 35$\%$. In a 30-day run during summer 2000,
corresponding to a total statistics of $\sim$ 10 millions $\phi$ decays,
they expect to reach a sensitivity ${\cal O}\sim 10^{-5}$ on the rare $\phi$ 
decays.

\section{Scalar Mesons}

A. Fariborz \cite{fariborz1}, in his talk on
 ``A Chiral Lagrangian Framework for Scalar
Mesons'', has drawn our attention on the important roles played by light
scalar mesons in low-energy QCD. Scalars are probes of the QCD vacuum and
are important from a phenomenological point of view, as they are
intermediate states in Goldstone boson interactions away from threshold, where
chiral perturbation theory is not applicable. Among the lowest-lying scalar
mesons ($ m<1$ GeV)  the $f_0(980)$ and $a_0(980)$ are rather well
established experimentally, the  $\sigma (560)$ or
$f_0(400-1200)$  has still uncertain mass and decay width  and  
the $\kappa(900)$ is not listed but mentioned in the PDG 2000.
It is known that a simple $q {\bar q}$ picture does not explain the
properties of these mesons, as already underlined by E. Oset in his talk.
\newline
The next-to-lowest scalars (1 GeV $< m <$ 2 GeV) are: $K_0^*(1430)$,
$a_0(1450)$, $f_0(1370)$, $f_0(1500)$, $f_J(1710)$ , and are all listed
in the PDG 2000. These states are generally believed to be closer to
$q {\bar q}$ objects, except for the $f_0(1500)$ which is a good candidate
for the lowest scalar glueball state. 
\newline
Within a non-linear chiral Lagrangian framework, developed by Fariborz and
collaborators \cite{fariborz2,fariborz3,fariborz4,fariborz5} different 
Goldstone boson interactions ($\pi\pi, \pi\eta, \pi
k$) away from threshold can be studied. In this approach, a description of
scattering amplitudes which are, to a good approximation, both crossing
symmetric and unitary, is possible. This leads to probing light scalar
mesons and extracting their unknown physical properties by fitting the
scattering amplitudes to experimental data.
\vskip 0.2cm
A big effort in studying the properties of the scalar mesons $f_0(980)$,
$a_0(980)$ in $\phi$ radiative decays is presently being done by the KLOE
experiment \cite{farilla1} as reported by A. Farilla \cite{rubin5} in her talk on 
``First KLOE results on scalar mesons''.  
Since the branching ratios of $\phi\to f_0\gamma$,$\phi\to a_0\gamma$
can range from 10$^{-4}$ to 10$^{-6}$, depending on the nature of
these particles, their accurate measurement at a $\phi$ factory, together
with  a precise determination of the shape of the
$\pi\pi$ mass spectrum and of the $\eta\pi$ mass spectrum can help in
establishing a realistic model for these particles. The KLOE detector is
suited for the detection of all-photon final states and final states with
photons and charged particles. During its first year of data taking in 1999 
the KLOE experiment,  which is presently running
at the DA$\Phi$NE collider in Frascati,
has collected 2.4 pb$^{-1}$ of integrated luminosity
corresponding to $\sim$8 millions $\phi$ decays.
This data sample has been used to study the radiative decays
$\phi\to f_0\gamma$,$\phi\to a_0\gamma$ \cite{farilla3}. The decay  
$\phi\to f_0\gamma\to\pi^{0}\pi^{0}\gamma$ has relevant background from 
$e^+e^-\to\omega\pi^0$. Other background processes are $\phi\to
\rho\pi^{0},\phi\to a_0 \gamma,\phi\to\eta\gamma$ with only photons in the
final state. A constrained kinematic fit and suitable topological cuts are
applied to reduce the background, with a final detection efficiency of
$\sim40\%$. A preliminary measurement gives  Br($\phi\to
f_0\gamma\to\pi^0\pi^0\gamma$)=
(0.81$\pm$0.09$_{stat}\pm$0.06$_{syst}$)$\cdot$10$^{-4}$.
The decay $\phi\to f_0\gamma\to\pi^{+}\pi^{-}\gamma$
has high background both from Initial State Radiation
and from Final State Radiation and interference with the latter is expected.
$\pi^{+}\pi^{-}\gamma$ events have been selected
with an overall efficiency of $\sim50\%$.
They obtain the upper limit: 
Br($\phi\to f_0\gamma\to\pi^+\pi^-\gamma$)$<$1.6$\cdot$10$^{-4}$ at 90\% C.L.
The decay $\phi\to a_0\gamma\to\eta\pi^0\gamma$ has similar background as
the $\phi\to f_0\gamma\to\pi^{0}\pi^{0}\gamma$ and again a 
constrained kinematic fit and suitable topological cuts are
applied to reduce the background, with a final detection efficiency of
$\sim23\%$. Their preliminary measurement gives  Br($\phi\to\eta\pi^0\gamma$)=
(0.77$\pm$0.15$_{stat}\pm$0.10$_{syst}$)$\cdot$10$^{-4}$.

\section{Form-factors and Polarizabilities}

There were four experimental and three
theoretical presentations in this section.  These fell into the
general areas of pion form-factors and polarizabilities,
measurements of anomalous amplitudes
$\pi \rightarrow \gamma \gamma$ and $\gamma \rightarrow 3\pi$, and
$J/\psi$ decays to $\phi (\omega) \pi^+ \pi^-$, $K^+ K^-$.

\subsection{Pion form-factors and polarizabilities}

T. Walcher\cite{Walcher} presented plans at Mainz for a new measurement
of the $\pi^+$ electromagnetic polarizability.
They plan to make a completely exclusive measurement of the
$\gamma p \rightarrow \pi^+ \gamma n$ reaction, where
all particles in the final state
are detected.  Preliminary results
from a test run at $E_{\gamma} =700$ MeV
show that the measured $\gamma p \rightarrow \pi^+ \gamma n$
cross sections are close to theoretical estimates, providing
confidence in the experimental apparatus and the theoretical
calculations that will be used to extract the polarizabilities.
D. Mack\cite{Mack} presented results from Jefferson Lab on the $\pi^+$
form factor, and plans for new measurements at higher $Q^2$.
Data were taken at W=1.95 GeV and
a Rosenbluth separation was made of the longitudinal and transverse
response functions.  A Regge analysis (VGL) of the longitudinal response
was used to extract the pion form factor.  The new data that go up to
$Q^2 \approx 2$ GeV$^2$ do not show evidence for
the $Q^2$ scaling that is expected in the PQCD regime. Higher
 $Q^2$ data are needed.

In theory J. Palomar \cite{palomar} presented calculations
of the pion and Kaon vector form factors using a
unitarization method to take into account final state
interaction corrections to the tree level amplitude from lowest
order CHPT.  This calculation describes very precisely the pion and
Kaon vector form factors and the p-wave $\pi \pi$ phase shifts
up to about s=1.44 GeV$^2$.  Agreement is much better with the two loop
CHPT pion vector form factor than with the one loop calculation.
A.~Radyushkin\cite{radyushkin} presented PQCD calculations of the $\pi^+$
and $\pi^0$ form factors, where it was shown how to combine
the exclusive QCD reaction mechanism with the others. This work shows that
there is some hope to reach perturbative QCD down to an energy level where
resummation methods as presented by Palomar and others are valid, thus holding
out hope for an eventual practically usable connection between
chiral dynamics and QCD.

\subsection{The anomalous reactions $\pi \rightarrow \gamma \gamma$ and
$\gamma \rightarrow 3\pi$}

There were two presentations that described measurements of
amplitudes dominated by the
Chiral anomaly. A. Gasparian\cite{gasparian} presented plans for a future
high precision measurement of the
$\pi^0$ lifetime at JLab, the PRIMEX experiment.
This experiment will measure the Primakoff production cross
section for $\pi^0$
on nuclear targets using 6 GeV incident tagged photons. A 1000+
element hybrid calorimeter that
consists of lead glass with a high resolution lead tungstate insert is
under construction for this experiment. The uncertainly in
measuring the $\pi^0$ lifetime is estimated to be less than 1.5\%.  There
are also
plans for measuring the $\pi^0$
form factor at low $Q^q$, and when higher energies are available, the
radiative widths and low $Q^q$ form factors of the $\eta$ and $\eta^\prime$.
R. Miskimen \cite{miskimen}
presented preliminary results on $\gamma \rightarrow 3\pi$
from an analysis of $\gamma p \rightarrow \pi^+ \pi^0 n$ data taken
on the CLAS detector at JLab.
The photon energy was approximately 2 GeV.
A Chew-Low analysis was used to
extract $F^{3\pi}$ from the cross sections over a range
in s from 18 to 38 $m_{\pi}^2$.
The results are in good agreement with a calculation by Holstein
 that
includes the effects of $\pi \pi$ FSI and, at low s, with several other
calculations\cite{Holstein}.
Analysis of the data at $s<18 m^2_{\pi}$
are continuing.

J. Oller \cite{Oller} presented calculations of
$J/\Psi \rightarrow \phi (\omega ) \pi^+ \pi^-$, $K^+ K^-$ where the
$\phi$ is treated as a spectator.  They obtained good agreement
with data up to $\sqrt{s} < 1.2$ GeV.  These decays are very
sensitive to OZI violating physics, supporting the statement that
the OZI rule is subjected to large corrections in the $0^{++}$
sector.

\section{Other}

\subsection{Chiral symmetry breaking and generalized CHPT}
\label{Other}

One of the main underlying assumptions in the way we normally perform
CHPT is that the quark-anti-quark vacuum-expectation-value is
of natural size, or equivalently, that the Gell-Mann-Oakes-Renner (GOR)
formula for the pion mass has only small corrections.
This assumption does not need to be made\cite{Fuchs} and it is in
fact surprisingly difficult to experimentally prove this assumption
even though a lot of qualitative evidence has been accumulated.
The measurement of the $\pi\pi$-threshold parameters has been promoted
as one of the places where there is a significant difference.
An introduction with lots of references can be found in the plenary
talk by Jan Stern in the previous Chiral Dynamics meeting\cite{Stern1}.
In this meeting there were references to this in the plenary talks
by Leutwyler\cite{Leutwyler} and Gasser\cite{Gasser}. In this working group
two new related developments in this area were discussed.

One of the questions in CHPT has always been how to determine the
large $N_c$ suppressed couplings and in particular $L_4^r$ and $L_6^r$.
In most fits to the $L_i^r$ these two are taken as input
using large $N_c$ argument and at best a check on the dependence
is done\cite{ABT}.
Moussallam\cite{Moussallam} presented how the scalar charge radius
of the pion and especially its strange scalar charge radius of the ion
can be used to determine these constants. In addition, sum rules
involving the scalar correlator
\begin{equation}
\Pi_S \sim \langle 0|T\left(\bar{s}s(x)(\bar{u}u+\bar{d}d)(0)
\right)|0\rangle
\end{equation}
can be used to put strong constraints on $L_6^r$.
He obtained positive values somewhat outside the large $N_c$-error bars
usually assumed. One consequence of these substantial violations of large $N_c$
or the OZI-rule is that the critical number of flavours above which
there is no spontaneous chiral symmetry breaking is expected to be
much less than $N_f=16.5$ where asymptotic freedom is no longer valid.

The same question was taken up by Stern\cite{Stern2}. Defining condensates
by the number of massless quark flavours, we look for deviations
from the GOR relation via 
the difference of $X(N_{\mbox{\tiny massless}})
= -2\hat{m}\langle \bar{u}u\rangle(N_{\mbox{\tiny massless}})/
(F_\pi^2 M_\pi^2)$ from 1. The conclusion is that $X(3)$
is rather suppressed unless $L_6^r(M_\rho)\approx-0.00026$.
The quantity $X(2)$ which is measured in $\pi\pi$-scattering threshold
parameters is not suppressed unless $r=m_s/\hat{m}<20$.
In fact, $X(2)-X(3)$ can be related to $\Pi_S(0)$ and must be positive.
This  means that $\bar l_3$ can be quite different from the usual
estimates since these involve $X(3)$. It is therefore quite possible
that standard CHPT is valid for two-flavours but not for three.
Notice however that Ref.\cite{ABT} find order 10\% corrections to $F_\pi$
and order 25 to 30\% to $M_\pi^2$ already accounting for
a 50\% correction in the denominator of $X(3)$.

\subsection{Large $N_c$ QCD and understanding hadronic parameters}

One of the subjects that was somewhat underrepresented at this meeting but
looming as a large background problem in many of the talks was the
connection between QCD and chiral dynamics at a higher than symmetry
considerations level. One of the more recent approaches here is
to combine large $N_c$-methods with meson dominance assumptions.
This approach was discussed shortly by Golterman\cite{Golterman}
where a curious relation between $F_\pi$ and $M_\rho$ was derived
from global duality in the large $N_c$ limit. This approach has grown
out of attempts at understanding the successes of chiral models like
the chiral quark model or the ENJL model. More references
are in Ref.\cite{Golterman}.

\subsection{CHPT methods at other energy scales}

The Higgs sector of the standard model is really just the linear sigma
model with some parts of the chiral symmetry group gauged. It is therefore
very tempting to simply replace it with a nonlinear sigma model. This
is basically the technicolor scenario. In its simplest version it
assumes a scale up of the QCD dynamics to the weak scale. This simplest version
is ruled out by the precision LEP experiments. However, simply replacing
the Higgs sector by the equivalent of the CHPT Lagrangian is not quite correct
as described by Nyffeler\cite{Nyffeler} in his contribution. There are
subtleties involved with gauge invariance and due to the presence of fermions
more terms are in principle possible.
In the light of this more general approach, LEP has not ruled out a
strongly interacting Higgs sector \`a la technicolor, but only its scaled
version from QCD. It is thus possible that future colliders will
serve us with a next level of Chiral Dynamics.

\section{$\mathbf \eta$}

The large $N_{\!c}$ limit of QCD \cite{kaiser1} is the background of the work presented by
R. Kaiser \cite{kaiser2} in his talk on ``Chiral perturbation theory and 
$1/N_{\!c}$-expansion'', an argument already discussed by H. Leutwyler in
his plenary talk\cite{Leutwyler}. 
Kaiser briefly reviews the effective theory that
describes the low energy properties of QCD with three light quarks and a
large number of colours, $N_{\!c}$, and then discusses the mechanisms that forbid
the Kaplan-Manohar transformation in this framework.
At large  $N_{\!c}$ the occurrence of a new energy scale $(M_{\eta'})$
leads to a more complicated structure of QCD. In the limit 
$N_{\!c}\to\infty$, the U(1)$_{\scriptscriptstyle A}$-anomaly is suppressed
and  this symmetry breaks down:
U(3)$_{\scriptscriptstyle R}\times$U(3)$_{\scriptscriptstyle L}\to U(3)_
{\scriptscriptstyle V}$. The formulation of the effective theory at large 
$N_{\!c}$ involves a simultaneous expansion in powers of momenta, quark
masses and $1/N_{\!c}$: $\delta$-expansion  \cite{kaiser4}.
The symmetry of Kaplan and Manohar is not realized at large
$N_{\!c}$. Within this framework quark mass ratios may be determined
unambiguously also at next-to-next-to leading order.
\vskip 0.2cm
F. Ambrosino \cite{rubin6} in his talk on 
''KLOE first results on $\eta$, $\eta'$'', reported about the preliminary
measurement of  $\phi\to\eta\gamma$, $\phi\to\eta'\gamma$  \cite{farilla3}
with the $\sim$8
millions $\phi$ decays detected by the KLOE experiment in the 1999 data
taking with 2.4$pb^{-1}$. The value of BR(
$\phi\to\eta'\gamma$) is a probe of the gluonium content of the $\eta'$
\cite{ambrosino2}  while the ratio 
$R=\frac{\Gamma(\phi\to\eta'\gamma)}{\Gamma(\phi\to\eta\gamma)}$ is
strictly related to the pseudoscalar mixing angle \cite{ambrosino3,ambrosino4}.
The decays  $\phi\to\eta\gamma$, $\phi\to\eta'\gamma$ have both been
studied in the final states $\pi^+\pi^-\gamma\gamma\gamma$ and $7\gamma$.
$\phi\to\eta\gamma$ decays are the main background for the rare
$\phi\to\eta'\gamma$ channel and at the same time they constitute a high
statistics control sample in the selection of  $\phi\to\eta'\gamma$ events.
Preliminary results show a perfect agreement between data and Monte
Carlo. A constrained kinematic fit, together with topological cuts, is used
to improve the S/B ratio. 
After the selection  $21\pm 4.6$
$\eta'$ events survive in the $\pi^+\pi^-\gamma\gamma\gamma$ final state and
$6^{+3.3}_{-2.2}$ in the $7 \gamma$ final state with less than one expected
background event at 90\% CL. This is the first observation of the
$\phi\to\eta'\gamma\to 7 \gamma$ decay chain. For the 
 $\pi^+\pi^-\gamma\gamma\gamma$ final state they obtain 
 $R=(7.1 \pm 1.6 ({\rm stat.}) \pm 0.3 ({\rm
  syst.}))\cdot 10^{-3}$ or BR($\phi\to\eta'\gamma$)=$ (8.9\pm 2 \pm
0.6)\cdot 10^{-5}$.
This result has the same level of accuracy of the
current world average.The high value of BR($\phi\to\eta'\gamma$) disfavours
models with large gluonium admixtures of the $\eta'$. 
Using formulas given in \cite{ambrosino4}, the value of $R$ is used 
to extract a mixing angle $\vartheta_P\simeq-19^{\circ}$. 
\vskip 0.2cm
$\eta$ decays are a powerful test of CHPT as underlined by B. Nefkens
\cite{nefkens1}
in his talk on ``New Tests of Chiral Perturbation Theory in $\eta$ Decays
using the Crystal Ball''. 
The G-parity violating $\eta\to 3\pi^0$ decay occurs primarily as a
consequence of the up-down quark mass difference, $m_u-m_d$.
Thus far CHPT has not
succeeded in accounting fully for the experimental $\eta\to3\pi$
decay rate. The role played by dynamical effects with terms ${\cal O}({p^6})$
may be explored by a precise measurement of the quadratic slope parameter
$\alpha$ in  $\eta\to3\pi^0$ decay which, according to present
theory
\cite{nefkens2}, is expected to be  
$\alpha = -(14~\mathrm{to}~7)\times 10^{-3}$.
The existing
experimental data from GAMS-2000 \cite{nefkens3} and Crystal Barrel
\cite{nefkens4} have rather low accuracy. The Crystal Ball \cite{nefkens5}
apparatus at the AGS has produced a 
 sample of $6\times10^6$ $\eta\to 3\pi^0$ in $\pi^-p\to\eta n$ near
 threshold and  a very pure subsample of
$1\times10^6$ $\eta\to3\pi^0$ decays, with a background $<1\%$, has been
selected  using a kinematic fit to the process
$\pi^-p\to\eta n\to3\pi^0n\to 6\gamma n.$ With this data the value 
$\alpha=-(33\pm 3_{\mathrm{stat}}\pm 6_{\mathrm{syst}})\times 10^{-3}$
which is in agreement with theory in the
sign but not in the value of $\alpha$.
Analogous measurement in  $\eta\to\pi^+\pi^-\pi^0$ decays is planned by the
WASA\cite{WASA} and KLOE collaborations:
a comparison of  $\alpha^{000}$ and  $\alpha^{+-0}$ in
an important test of isospin invariance. The Crystal Ball collaboration
has also performed preliminary measurements of $\eta$ rare decays such as
$\eta\to\pi^0\gamma\gamma$, $\eta\to3\gamma$ and
$\eta\to\pi^0\pi^0\gamma\gamma$.

\section{Conclusions}

From the breadth of topics covered in this working group it is obvious that
chiral dynamics in the Goldstone sector is a very active field with 
progress both on the experimental and theoretical front.
As is also obvious, enough puzzles and challenges remain to keep all of
us busy till the next Chiral Dynamics meeting.

\section*{Acknowledgements}

We would like to thank the organizers for the pleasant atmosphere they created
at the meeting and for the smooth running of the entire organization.
The work of JB is partially supported the EU TMR Network
EURODAPHNE (Contract No. ERBFMX-CT98-0169).


\begin{thebibliography}{99}

\bibitem{Nakano} T.~Nakano, these proceedings;
S.~Adler {\it et al.}, hep-ex/0007021.

\bibitem{Schmidt} S.~Schmidt, these proceedings.

\bibitem{Arenton} M.~Arenton, these proceedings

\bibitem{Isidori} G.~Isidori, plenary talk, these proceedings.

\bibitem{BEP}
J.~Bijnens,
 G.~Ecker and A.~Pich,
Phys.\ Lett.\  {\bf B286} (1992) 341
[hep-ph/9205210];
G.~Ecker,
H.~Neufeld and A.~Pich,
Nucl.\ Phys.\  {\bf B413} (1994) 321
[hep-ph/9307285];
G.~D'Ambrosio {\it et al.},
Z.\ Phys.\  {\bf C76} (1997) 301
[hep-ph/9612412].

\bibitem{NA48KS}
A.~Lai {\it et al.}  [NA48 Collaboration],
``A new measurement of the branching ratio of $K_S\to\gamma\gamma$,''
CERN-EP-2000-122.

\bibitem{KSgg}
G.~D'Ambrosio and D.~Espriu,
Phys.\ Lett.\  {\bf B175} (1986) 237;
J.~L.~Goity,
Z.\ Phys.\  {\bf C34} (1987) 341.

\bibitem{KLpgg}
G.~Ecker, A.~Pich and E.~de Rafael,
Phys.\ Lett.\  {\bf B189} (1987) 363.

\bibitem{eeee}
G.~D'Ambrosio and J.~Portoles,
Nucl.\ Phys.\  {\bf B533} (1998) 494
[hep-ph/9711211] and references therein.

\bibitem{kppee}
A.~Alav-Harati {\it et al.}  [KTeV Collaboration],
Phys.\ Rev.\ Lett.\  {\bf 84} (2000) 408
[hep-ex/9908020].

\bibitem{KTeV}
A.~Alavi-Harati {\it et al.}  [KTeV Collaboration],
Phys.\ Rev.\ Lett.\  {\bf 83} (1999) 22
[hep-ex/9905060].

\bibitem{Cenci} P.~Cenci, these proceedings

\bibitem{NA48}
V.~Fanti {\it et al.}  [NA48 Collaboration],
Phys.\ Lett.\  {\bf B465} (1999) 335
[hep-ex/9909022].

\bibitem{GilmanWise}
F.~J.~Gilman and M.~B.~Wise,
Phys.\ Lett.\  {\bf B83} (1979) 83;
Phys.\ Rev.\  {\bf D20} (1979) 2392.

\bibitem{Franzini} P.~Franzini, plenary talk, these proceedings.

\bibitem{GoltermanPlenary} M.~Golterman, plenary talk, these proceedings.

\bibitem{Bijnenseps} J.~Bijnens and J.~Prades, these proceedings.

\bibitem{Benasque}
J.~Bijnens,
``Weak interactions of light flavors,''
hep-ph/0010265.

\bibitem{Cirigliano} V.~Cirigliano, these proceedings;
V.~Cirigliano, J.~F.~Donoghue and E.~Golowich,
hep-ph/0008290 and references therein.

\bibitem{EMpenguin}
J.~Bijnens and M.~B.~Wise,
Phys.\ Lett.\  {\bf B137} (1984) 245;
J.~M.~Flynn and L.~Randall,
Phys.\ Lett.\  {\bf B224} (1989) 221.

\bibitem{EIMNP}
E.~de Rafael,
Nucl.\ Phys.\ Proc.\ Suppl.\  {\bf 7A} (1989) 1;
G.~Ecker {\it et al.},
hep-ph/0006172.

\bibitem{Zeller} M.~Zeller, plenary talk, these proceedings.

\bibitem{PT}
G.~Colangelo,
M.~Knecht and J.~Stern,
Phys.\ Lett.\  {\bf B336} (1994) 543
[hep-ph/9406211].

\bibitem{AB}
G.~Amoros and J.~Bijnens,
J.\ Phys.\ G {\bf G25} (1999) 1607
[hep-ph/9902463].

\bibitem{Bijnenskl4} G.~Amoros, J~Bijnens and P.~Talavera, these proceedings.

\bibitem{KL4}
J.~Bijnens,
Nucl.\ Phys.\  {\bf B337} (1990) 635;
C.~Riggenbach {\it et al.},
Phys.\ Rev.\  {\bf D43} (1991) 127;
J.~Bijnens, G.~Colangelo and J.~Gasser,
Nucl.\ Phys.\  {\bf B427} (1994) 427
[hep-ph/9403390].

\bibitem{ABT}
G.~Amoros, J.~Bijnens and P.~Talavera,
Nucl.\ Phys.\  {\bf B568} (2000) 319
[hep-ph/9907264];
Phys.\ Lett.\  {\bf B480} (2000) 71
[hep-ph/9912398];
Nucl.\ Phys.\  {\bf B585} (2000) 293
[hep-ph/0003258].

\bibitem{Rosselet}
L.~Rosselet {\it et al.},
Phys.\ Rev.\  {\bf D15} (1977) 574.

\bibitem{Colangelo}
G.~Colangelo, plenary talk, these proceedings.

\bibitem{Leutwyler} H.~Leutwyler, plenary talk, these proceedings.

\bibitem{Roynew}
B.~Ananthanarayan {\it et al.},
hep-ph/0005297.
G.~Colangelo, J.~Gasser and H.~Leutwyler,
Phys.\ Lett.\  {\bf B488} (2000) 261
[hep-ph/0007112].

\bibitem{Royold}
M.~Knecht {\it et al.},
Nucl.\ Phys.\  {\bf B471} (1996) 445
[hep-ph/9512404].

\bibitem{Gasser} J.~Gasser, plenary talk, these proceedings.

\bibitem{Rusetsky} A.~Rusetsky, plenary talk, these proceedings;
V.~Antonelli {\it et al.},
hep-ph/0003118;
A.~Gall {\it et al.},
Phys.\ Lett.\  {\bf B462} (1999) 335
[hep-ph/9905309].

\bibitem{Schacher} J.~Schacher, these proceedings.

\bibitem{Saborges}
I. Cavalcante and J. S\'a Borges, these proceedings.

\bibitem{pelaez1} J. Pelaez, these proceedings.

\bibitem{pelaez2} J.A. Oller, E. Oset and J.R. Pel\'aez, Phys. Rev. {\bf D62}
(2000) 114017 [hep-ph/9911297].

\bibitem{oset1} E. Oset, these proceedings.

\bibitem{oset2} J. A. Oller, E. Oset and J. R. Pel\'aez, Phys.\ Rev.\ Lett.\
  {\bf 80}, 3452 (1998); 
Phys.\ Rev.\ D {\bf 59} (1999) 074001;
 Erratum-ibid. D {\bf 60} (1999) 099906.

\bibitem{oset3} F. Guerrero and J. A. Oller, Nucl.\ Phys.\ B {\bf 537},
(1999) 459.

\bibitem{oset4} J. A. Oller and E. Oset,
 Phys.\ Rev.\ D {\bf 60}, 074023 (1999).

\bibitem{rubin1} P. Rubin, these proceedings.

\bibitem{rubin2} M.N. Achasov {\it et al.}, Phys.Lett. B479 (2000) 53;
M.N. Achasov {\it et al.}, hep-ex/0005017

\bibitem{rubin3} R.R. Akhmetshin {\it et al.}, Phys.Lett. B462 (1999) 371;
 R.R. Akhmetshin {\it et al.}, Phys.Lett. B462 (1999) 380

\bibitem{rubin5} A. Farilla, these proceedings.

\bibitem{rubin6} F. Ambrosino, these proceedings.

\bibitem{fariborz1} A. Fariborz, these proceedings.

\bibitem{fariborz2} F. Sannino and J. Schechter,
\Journal{\PRD}{52}{96}{1995}, and
M. Harada,
F. Sannino and J. Schechter,
\Journal{\PRD}{54}{1991}{1996}; \Journal{\PRL}{78}{1603}{1997}.

\bibitem{fariborz3}D. Black {\it et al.},
\Journal{\PRD}{58}{054012}{1998}; \Journal{\PRD}{59}{074026}{1999}.

\bibitem{fariborz4}A.H. Fariborz and J. Schechter, \Journal{\PRD}{60}{034002}
{1999}.

\bibitem{fariborz5}D. Black, {\it et al.},%
A.H. Fariborz and J. Schechter,
\Journal{\PRD}{61}{074001}{2000}; \Journal{\PRD}{61}{074030}{2000}.

\bibitem{farilla1} S. Bertolucci for the KLOE collaboration, hep-ex/0002030.

\bibitem{farilla3} The KLOE collaboration, Contributed paper to ICHEP2000
  (Osaka), hep-ex/0006036. 

\bibitem{Walcher} T.~Walcher, these proceedings.

\bibitem{Mack} D.~Mack, these proceedings.

\bibitem{palomar} J.E. Palomar, E. Oset and J. Oller, these proceedings.

\bibitem{radyushkin}
A.~Radyushkin, these proceedings;
A.~P.~Bakulev, A.~V.~Radyushkin and N.~G.~Stefanis,
Phys.\ Rev.\  {\bf D62} (2000) 113001
[hep-ph/0005085].

\bibitem{gasparian} A. Gasparian, these proceedings.

\bibitem{miskimen} R. Miskimen and B. Asavapibhop, these proceedings.

\bibitem{Holstein} 
B.~R.~Holstein,
Phys.\ Rev.\  {\bf D53} (1996) 4099
[hep-ph/9512338];
J.~Bijnens, A.~Bramon and F.~Cornet,
Phys.\ Lett.\  {\bf B237} (1990) 488.


\bibitem{Oller} J. Oller and U.-G. Mei\ss{}ner, these proceedings.

\bibitem{Fuchs}
N.~H.~Fuchs, H.~Sazdjian and J.~Stern,
Phys.\ Lett.\  {\bf B269} (1991) 183.

\bibitem{Stern1}
J.~Stern,
``Light quark masses and condensates in QCD,''
in Mainz 1997, Chiral dynamics: Theory and experiment, eds.
A.~Bernstein, D.~Drechsler and Th. Walcher
[hep-ph/9712438].

\bibitem{Moussallam}
B.~Moussallam, these proceedings;
JHEP {\bf 0008} (2000) 005
[hep-ph/0005245].

\bibitem{Stern2}
S.~Descotes and J.~Stern, these proceedings;
S.~Descotes and J.~Stern,
Phys.\ Lett.\  {\bf B488} (2000) 274
[hep-ph/0007082];
S.~Descotes, L.~Girlanda and J.~Stern,
JHEP {\bf 0001} (2000) 041
[hep-ph/9910537].

\bibitem{Golterman}
M.~Golterman and S.~Peris, these proceedings;
M.~F.~Golterman and S.~Peris,
Phys.\ Rev.\  {\bf D61} (2000) 034018
[hep-ph/9908252];
S.~Peris, M.~Perrottet and E.~de Rafael,
JHEP {\bf 9805} (1998) 011
[hep-ph/9805442].


\bibitem{Nyffeler}
A.~Nyffeler, these proceedings;
A.~Nyffeler and A.~Schenk,
Phys.\ Rev.\  {\bf D62} (2000) 113006
[hep-ph/9907294].

\bibitem{kaiser1} R. Kaiser and H. Leutwyler, hep-ph/0007101 and references
therein.

\bibitem{kaiser2} R. Kaiser, these proceedings.

\bibitem{kaiser4} H. Leutwyler, Phys.\ Lett.\  {\bf B374}, 163 (1996)
[hep-ph/9601234].

\bibitem{ambrosino2} N. G. Deshpande, G. Eilam, \Journal{\PRD}{25}{270}{1982}

\bibitem{ambrosino3} J. L. Rosner, \Journal{\PRD}{27}{1101}{1983}

\bibitem{ambrosino4}  A. Bramon {\em et al.}, {\em Eur. Phys. J.} C {\bf 7},
  271 (1999)
\bibitem{nefkens1} B.M.K. Nefkens and S.~Prakhov, these proceedings.

\bibitem{nefkens2} J.~Kambor,
C.~Wiesendanger and D.~Wyler,
Nucl.\ Phys.\  {\bf B465} (1996) 215;
A.~V.~Anisovich and H.~Leutwyler,
Phys.\ Lett.\  {\bf B375} (1996) 335.

\bibitem{nefkens3} The GAMS-2000 Collaboration, D. Alde {\it et al.},
  \Journal{\ZPC}{25}{225}{1984}

\bibitem{nefkens4} The Crystal Barrel Col., A. Abele {\it et al.},
 \Journal{\PLB}{417}{193}{1998}.

\bibitem{nefkens5} The Crystal Ball Col., S.Prakhov {\it et al.},
 \Journal{\PRL}{84}{4802}{2000}.

\bibitem{WASA} H.~Cal\'en {\it et al.},
Nucl.\ Instrum.\ Meth.\  {\bf A379} (1996) 57.
\end{thebibliography}
\end{document}